\documentclass[11pt]{article}
\usepackage[hmargin=1in, vmargin=1in]{geometry} 
\usepackage{amsmath}
\usepackage{amssymb}
\usepackage{amsthm}
\usepackage{bbm}
\usepackage{graphicx}
\usepackage{natbib}
\usepackage[multiple]{footmisc}
\begin{document}
\newcommand{\indicator}[1]{\mathbbm{1}\left[ {#1} \right] }
\newcommand*\wrapletters[1]{\wr@pletters#1\@nil}
\def\wr@pletters#1#2\@nil{#1\allowbreak\if&#2&\else\wr@pletters#2\@nil\fi}


\title{Spectral gene set enrichment (SGSE)}

\author{H. Robert Frost
\footnote{Institute for Quantitative Biomedical Sciences, Geisel School of Medicine, Lebanon, NH 03756} 
\footnote{Section of Biostatistics and Epidemiology, Department of Community and Family Medicine, Geisel School of Medicine, Lebanon, NH 03756} 
\footnote{Department of Genetics, Dartmouth College, Hanover, NH 03755}, Zhigang Li$^{*\dagger}$ and Jason H. Moore\,$^{*\dagger\ddag}$}

\maketitle

\begin{abstract}

\noindent \textbf{Motivation:} 
Gene set testing is typically performed in a supervised context to quantify the association between groups of genes and a clinical phenotype. In many cases, however, a gene set-based interpretation of genomic data is desired in the absence of a  phenotype variable. Although methods exist for unsupervised gene set testing, they predominantly compute enrichment relative to clusters of the genomic variables with performance strongly dependent on the clustering algorithm and number of clusters.
\noindent \textbf{Results:} 
We propose a novel method, spectral gene set enrichment (SGSE), for unsupervised competitive testing of the association between gene sets and empirical data sources.
SGSE first computes the statistical association between gene sets and principal components (PCs) using our principal component gene set enrichment (PCGSE) method. The overall statistical association between each gene set and the spectral structure of the data is then computed by combining the PC-level p-values using the weighted Z-method with weights set to the PC variance scaled by Tracey-Widom test p-values. 
Using simulated data, we show that the SGSE algorithm can accurately recover spectral features from noisy data.
To illustrate the utility of our method on real data, we demonstrate the superior performance of the SGSE method relative to standard cluster-based techniques for testing the association between MSigDB gene sets and the variance structure of microarray gene expression data.
\noindent \textbf{Availability:} http://cran.r-project.org/web/packages/PCGSE/index.html
\noindent \textbf{Contact:} rob.frost@dartmouth.edu or jason.h.moore@dartmouth.edu
\end{abstract}

\section{Introduction}

Gene set testing has become an indispensable tool for the analysis and interpretation of high dimensional genomic data, including measures of DNA sequence variation, DNA methylation, RNA expression and protein abundance \citep{Khatri:2012fk, Hung:2012kx}. By focusing on the collective effect of biologically meaningful groups of genomic variables, rather than just the marginal effect of individual variables,
gene set testing methods can significantly improve statistical power, replication of results and biological interpretation. Because of these benefits, significant effort has been devoted over the last decade to building large repositories of functional gene sets \citep{ashburner_gene_2000, kanehisa_kegg:_2000,Liberzon:2011uq}, creating methods for refining and customizing these gene set collections \citep{alterovitz_ontology_2010, Davis:2010dq, Frost:2014fk} and developing effective statistical techniques for gene set testing \citep{subramanian_gene_2005, Efron:2007uq, Barry2008, Frost:2012zr, Wu:2012fk, Zhou:2013ys}.

Gene set testing is normally used to quantify the association between functional groups of genomic variables and a clinical phenotype, e.g., cancer case/control status. Many important use cases exist, however, where a gene set-based interpretation of genomic data is desired in the absence of a phenotype variable, e.g., case-only data collections. For such unsupervised applications, the standard approach for gene set testing involves the computation of the association between gene sets and a categorical variable defined by disjoint clusters of genomic variables. Such methods typically compute the association between each gene set and the variable clustering using either information theoretic measures \citep{Gibbons:2002fk, Steuer:2006vn} or contingency table-based statistical tests which incorrectly assume independence among the genomic variables \citep{Robinson:2002fk, Toronen:2004uq, Freudenberg:2009ys}. 

Although these techniques provide a measure of gene set enrichment for a given clustering of genomic data, the motivation for most methods is cluster evaluation rather than unsupervised biological interpretation. Cluster-based gene set enrichment results are strongly dependent on the clustering method employed and the number of computed clusters. This sensitivity to the clustering method and number of clusters makes these methods very useful for clustering evaluation but unreliable as general measures of unsupervised gene set enrichment. Specifically, since these methods advocate the use gene set enrichment results to select the clustering method and number of clusters, instead of often unreliable metrics such as the gap statistic \citep{Tibshirani:2001fk} or average silhouette width \citep{Kaufman:2005vn}, it is unclear what clustering method or number of clusters should be used if the goal is unbiased gene set testing.	 

An alternative approach for unsupervised gene set testing with many similarities to cluster-based methods is gene set enrichment of gene networks.
This approach typically involves the computation of a network from a genomic dataset with network nodes represented by genomic variables, e.g., a co-expression network for gene expression data \citep{Zhao:2010fk}, a community detection algorithm is then used to decompose the network nodes into distinct groups and, finally, gene set testing is performed relative to each community or all communities. If the network communities are treated like gene clusters, the same information theoretic or contingency table-based methods employed for cluster-based enrichment can be used to calculate the association between gene sets and network communities. Approaches have also been developed that directly leverage the network structure to test for the association between gene sets and single network nodes \citep{Wolfe:2005ly} or groups of nodes \citep{Glaab:2012kx}. Similar to cluster-based approaches, gene set enrichment of networks is highly dependent on the method used to build the network from genomic data and algorithms employed for community detection. 

Methods have also been developed to test the association between gene sets and latent variables computed from genomic data sets via techniques such as principal component analysis (PCA) or independent component analysis (ICA). Most of these methods test for the association with just a single latent variable and employ an anti-conservative contingency-table based test on a dichotomized version of the loadings for the latent variable \citep{Lee:2003bh, roden_mining_2006, Yao:2012qf}. An exception is our recently developed principal component gene set enrichment (PCGSE) method \citep{Frost:PCGSE_arXiv_2014} that performs competitive gene set testing relative to each PC using a statistical test that adjusts for correlation among gene set members. Similar to single cluster gene set testing methods, methods that perform gene set testing relative to a single latent variable can only provide an interpretation for a portion of a genomic data set. To test for the association between gene sets and a collection of latent variables representative of the entire data set, matrix correlation methods \citep{jolliffe2002principal, Ramsay:1984fk} have been employed, however, such methods are dependent on the number of latent variables included in the test and can only be used for self-contained gene set testing \citep{Goeman:2007pr} ($Q_2$ in the terminology of Tian et al. \citep{Tian:2005zr}).

Effective methods do not currently exist for unsupervised gene set testing against a competitive null hypothesis that are independent of specific cluster analysis or network analysis approaches.  To address this shortcoming, we have developed spectral gene set enrichment (SGSE), an approach for unsupervised competitive testing of the association between gene sets and empirical data sources independent of cluster or network analysis. The SGSE method first computes the statistical association between gene sets and principal components (PCs) using our principal component gene set enrichment (PCGSE) method. The overall statistical association between each gene set and the spectral structure of the data is then computed by combining the PC-level p-values using the weighted Z-method with weights set to the PC variance scaled by lower-tailed p-value computed for the PC variance according to the Tracey-Widom distribution. Although described in the context of gene sets and genomic data, the SGSE method can be used to compute the statistical association between any collection of variable groups and the spectral structure of any empirical data set. To facilitate use of the SGSE method by other researchers, we have included an implementation of the algorithm in the PCGSE R package, which is available from the CRAN repository. Using simulated gene expression data and simulated gene sets, we show that the SGSE method can accurately recover known spectral features from noisy data, features that are undetectable using cluster-based approaches. To illustrate the utility of our method on real genomic data, we compare the performance of the SGSE method and a cluster-based technique on testing the association between MSigDB gene sets and the spectra of two cancer microarray gene expression data sets. 

\section{Methods}

\subsection{SGSE inputs}
Similar to our principal component gene set enrichment (PCGSE) method \citep{Frost:PCGSE_arXiv_2014}, the SGSE method takes as input both an $n \times p$ genomic data matrix $\mathbf{X}$ quantifying $p$ genomic variables under $n$ experimental conditions and an $f \times p$ binary annotation matrix $\mathbf{A}$ that specifies the association between the $p$ genomic variables and $f$ functional categories.   

The genomic data held in $\mathbf{X}$, e.g., mRNA expression levels, will be modeled as a sample of $n$ independent observations from a $p$-dimensional random vector $\mathbf{x}$.  It is assumed that any desired transformations on $\mathbf{X}$ have been performed and that missing values have been imputed or removed. Although the SGSE method is robust to departures from multivariate normality, as discussed in Section \ref{sec:rmt} below, it will be assumed that $\mathbf{x} \sim MVN(\boldsymbol{\mu}, \boldsymbol{\Sigma})$ with correlation matrix $\mathbf{P}$. This distributional assumption is usually well justified since sources of genomic data, especially gene expression data, typically follow a multivariate normal distribution after appropriate transformations. 

The rows of $\mathbf{A}$ represent $f$ biological categories, e.g., KEGG pathways or GO categories, and the elements $a_{i,j}$ hold indicator variables whose value depends on whether an annotation exists between the function $i$ and genomic variable $j$.  

\subsection{SGSE algorithm}

Enrichment of the gene sets defined by $\mathbf{A}$ relative to the spectra of $\mathbf{X}$ is performed using the following steps, which are explained in detail in sections \ref{sec:pca} thru \ref{sec:liptaks} below. 
\begin{enumerate}
\item Perform PCA on $\mathbf{X}$.
\item Determine $q$, the number of PCs used to represent the spectra of $\mathbf{X}$.
\item For all $q$ PCs, use the PCGSE method to compute the statistical significance of the association between each PC and each of the $f$ gene sets defined by $\mathbf{A}$ according to a competitive null hypothesis.
\item Compute the statistical significance of the association between each of the $f$ gene sets and the spectra of $\mathbf{X}$ using the weighted Z-method on the $q$ PCGSE p-values with weights based on the PC variances optionally scaled according to PC statistical significance.
\end{enumerate}

\subsection{PCA for SGSE}\label{sec:pca}

Because PCs are not invariant under scaling of the data \citep{jolliffe2002principal}, the PCA solution for SGSE is computed on a mean centered and standardized version of $\mathbf{X}$, $\mathbf{\tilde{X}}$. The PC loading vectors and variances of $\mathbf{\tilde{X}}$ are thus the eigenvectors and eigenvalues of: 
\begin{equation}\label{eq:S}
	\mathbf{S} = \frac{1}{n-1} \mathbf{\tilde{X}}^T\mathbf{\tilde{X}}
\end{equation}
\noindent The spectral decomposition of $\mathbf{S}$ is defined as:
\begin{equation} \label{eqn:spectral_decomp}
	\mathbf{S} = \sum_{i=1}^{r_{\mathbf{\tilde{X}}}} \lambda_i v_i v_i^T
\end{equation}
\noindent where $r_{\mathbf{\tilde{X}}}$ is the rank of $\mathbf{\tilde{X}}$, $\lambda_i$ is the $i^{th}$ eigenvalue of $\mathbf{S}$ and the variance of the $i^{th}$ PC of $\mathbf{\tilde{X}}$, $v_i$ is the $i^{th}$ eigenvector of $\mathbf{S}$ and the loadings for the $i^{th}$ PC of $\mathbf{\tilde{X}}$ and $\mathbf{\tilde{X}} v_i$ is the $i^{th}$ PC. It is assumed that the eigenvalues are sorted in decreasing order: $\lambda_1 \geq \lambda_2 \geq ... \geq \lambda_{r_{\mathbf{\tilde{X}}}}$. Because $\mathbf{x} \sim$ MVN, $(n-1)\mathbf{S}$ is approximately Wishart distributed:
\begin{equation}	
	(n-1)\mathbf{S} = \mathbf{\tilde{X}^T\tilde{X}} \sim W(n, \mathbf{P})
\end{equation}
\noindent Similar to PCGSE, the PCA solution for SGSE realized via the singular value decomposition (SVD) of a, $\mathbf{\tilde{X}} = \mathbf{U} \mathbf{E} \mathbf{V}^T$, \noindent where the columns of $\mathbf{V}$ represent the PC loading vectors, the entries in the diagonal matrix $\mathbf{E}$ are proportional to the square roots of the PC variances and the columns of $\mathbf{U} \mathbf{E}$ are the PCs.

\subsection{PC statistical significance}\label{sec:rmt}

Random matrix theory (RMT) methods provide useful distributional results for the bulk and extreme eigennvalues of matrices with Wishart distributions \citep{johnstone_2001, izenman_rmt, Johnstone:2009fk}. As outlined by \cite{johnstone_2001}, the principal eigenvalue of a sample covariance matrix with a \textit{white} Wishart distribution, where \textit{white} implies that $\mathbf{\Sigma} = \mathbf{I}$, tends to a distribution described by a \textit{Tracey-Widom} law of order 1 \citep{Tracy:1994fk}. Specifically, if $n, p \to \infty, n/p \to \eta \geq 1$, then the distribution of the rescaled principal eigenvalue:
\begin{equation} \label{eq:tw_dist}
	\frac{\lambda_1 - \mu(p, n)}{\sigma(p, n)} 
 \end{equation}
tends to a \textit{Tracey-Widom} law of order 1, where $\mu(p, n) = \frac{(\sqrt{n-1} + \sqrt{p})^2}{n}$ and $\sigma(p, n) = \frac{\sqrt{n-1} + \sqrt{p}}{n} \left(\frac{1}{\sqrt{n-1}} +\frac{1}{\sqrt{p}} \right)^{1/3}$.
\noindent For $p > n$, the \textit{Tracey-Widom} distribution still holds with $p$ and $n$ simply reversed in the $\mu(p, n)$ and $\sigma(p, n)$ parameter definitions. Although an asymptotic result, this distribution was found to hold well even for $p$ and $n$ values as small as $p=20$ and $n=5$ \citep{johnstone_2001, izenman_rmt}. It also holds well even when the underlying distribution of the elements of $\mathbf{\tilde{X}}$ is not normal \citep{Soshnikov02anote}.

In cases where $\mathbf{\Sigma}$ has $q$ variances greater than 1, i.e., a spiked covariance model, \cite{johnstone_2001} demonstrated that this distribution approximates the distribution of the $(q+1)^{th}$ eigenvalue but with slightly heavier tails. This result can therefore be used to compute a conservative statistical significance for all of the PCs of $\mathbf{\tilde{X}}$ based on the associated eigenvalues under a null hypothesis of uncorrelated MVN data (e.g., use of PCA for population genetics \citep{Patterson:2006uq}). Specifically, the statistical significance of PC $i$ can be determined by first computing a \textit{Tracey-Widom} distributed statistic, $tw_i$, for the eigenvalue, $\lambda_i$, associated with the PC using equation \eqref{eq:tw_dist}:
\begin{equation} \label{eq:tw_dist_i}
	tw_i = \frac{\lambda_i - \mu(p-i+1, n)}{\sigma(p-i+1, n)} 
 \end{equation}
Under the $H_0$ that the data is a sample from a MVN distribution with no pair-wise correlation among the individual variables, the p-value for PC $i$ is then computed as the probability of a \textit{Tracey-Widom} law of order 1 statistic more extreme than $tw_i$:
\begin{equation} \label{eq:tw_pval}
	\text{p-value}_{\text{PC}_i} = 1-F_{\text{TW}}(tw_i)
 \end{equation}
\noindent where $F_{TW}()$ is the cumulative distribution function of a \textit{Tracey-Widom} law of order 1 random variable. This probability can be computed using either numerical lookup tables such as those supported by the RMTstat R package or via the Gamma approximation to the \textit{Tracey-Widom} distribution detailed in \cite{2012arXiv1209.3394C}. The SGSE method currently uses the Gamma approximation for more accurate coverage of the tails of the distribution.

\subsection{Number of PCs used to represent data}\label{sec:num_pcs}

The SGSE method supports three options for determining $q$, the number of PCs used to represent the spectra of $\mathbf{X}$:
\begin{enumerate}
\item All PCs with non-zero variance: $q = \max_{i} \lambda_i > 0 = r_{\mathbf{\tilde{X}}}$
\item All statistically significant PCs at a specific $\alpha$ level where statistical significance of a given PC $i$ is determined according \eqref{eq:tw_pval}.
\item A specified number, $q^{*}$, with the constraint that $q^{*}$ cannot be greater than the number of PCs with non-zero variance: $q = q^{*}, \text{s.t. }  q^{*} \leq r_{\mathbf{\tilde{X}}}$. If specified, $q^{*}$ will typically be set to a small number, e.g., 1 or 2, to minimize computational cost of the SGSE algorithm.
\end{enumerate}

\subsection{PCGSE for SGSE}\label{sec:pcgse_for_sgse}

Our PCGSE method \citep{Frost:PCGSE_arXiv_2014} is used to compute the statistical significance of the association between each of the $f$ gene sets defined in $\mathbf{A}$ and each of the first $q$ PCs of $\mathbf{\tilde{X}}$ where $q$ is determined using one of the three methods detailed in Section \ref{sec:num_pcs} above. Let the p-value computed via PCGSE for PC $i$ and gene set $j$ be represented using the notation $\text{p-value}_{\text{PC}_i, \text{gs}_j}$. 

Although any supported PCGSE options can be used with SGSE, by default, the SGSE method executes PCGSE using the Fisher-transformed Pearson correlation coefficient between each variable and each PC as the gene-level test statistic and the correlation-adjusted standardized mean difference statistic as the gene set test statistic with statistical significance of the gene set test statistic under a competitive $H_0$ computed using a two-sided t-test as detailed in \cite{Frost:PCGSE_arXiv_2014}. 

\subsection{Combined significance of PCGSE p-values}\label{sec:liptaks}

For each of the $f$ gene sets, the p-values computed via PCGSE for the $q$ selected PCs of $\mathbf{\tilde{X}}$ are combined using the weighted Z-method, a generalization of the untransformed Z-transform test \citep{Whitlock:2005uq, Won:2009fk}. The weighted Z-method combines Z-statistics generated for each of multiple independent p-values using weights specific to each p-value.  This approach for combining p-values is justified for SGSE under the assumption of multivariate normality for $\mathbf{x}$ making both the uncorrelated PCs of $\mathbf{\tilde{X}}$, and the p-values generated by PCGSE with respect to those PCs, independent. If $\mathbf{x}$ is significantly non-Gaussian, then the PC-specific p-values will be dependent and techniques such as Kost's method \citep{Kost2002183} must be employed instead of the weighted Z-method. In the context of SGSE, a weighed Z-statistic is generated for each of the $f$ gene sets as follows:
\begin{equation} \label{eq:liptak}
	Z_{\text{gs}_j} = \frac{\sum_{i=1}^{q} w_{i} \Phi^{-1} (1-\text{p-value}_{\text{PC}_i, \text{gs}_j})}{\sqrt{\sum_{i=1}^{q} w_i^2 }}
 \end{equation}
\noindent where $w_i$ is a weight specific to PC $i$ of $\mathbf{\tilde{X}}$ and $\Phi^{-1}()$ is the inverse standard normal CDF.  Two options are supported for determining the PC-specific weights, $w_i$:
	\begin{enumerate}
	\item The weight is set to the variance of each PC: $w_i = \lambda_i$
	\item The weight is set to the variance of each PC scaled by the lower-tailed p-value computed for the PC variance according to the \textit{Tracey-Widom} distribution as detailed in Section \ref{sec:rmt}: $w_i = (1-\text{p-value}_{\text{PC}_i}) \lambda_i = F_{\text{TW}}(tw_i) \lambda_i$
	\end{enumerate}
The overall p-value representing the statistical significance of the association between gene set $j$ and the spectra of $\mathbf{X}$ is then computed using a one-sided z-test on $Z_{\text{gs}_i}$: 
\begin{equation} \label{eq:sgse_sig}
	\text{p-value}_{\text{gs}_j} = 1-\Phi(Z_{\text{gs}_j})
 \end{equation}
\subsection{SGSE evaluation}

\subsubsection{Benchmark cluster-based gene set testing method}\label{sec:cluster_method}
To support comparative evaluation of the SGSE method, we implemented a cluster-based gene set testing method that is representative of a large number of existing cluster and network-based gene set testing methods. Our benchmark cluster-based method computes the statistical significance of the association between gene sets and a data set as follows:
\begin{enumerate}
\item Cluster the $p$ genomic variables in $\mathbf{\tilde{X}}$ using k-means clustering with the Hartigan and Wong algorithm \citep{hartigan1979}, 5 restarts and k set according to the global maximum of the gap statistic \citep{Tibshirani:2001fk} as computed using the \textit{clusGap()} function in the \textit{cluster} R package \citep{Maechler:2014fk} with the number of bootstrap resamples defaulting to 100.
\item Compute the statistical significance of the association between each of the $f$ gene sets defined in $\mathbf{A}$ and the k-means clustering using Pearson's $\chi^2$ test of independence on a $2 \times k$ contingency table whose first row holds the counts of gene set members in each of the k clusters and whose second row holds the total size of each of the k clusters.
\end{enumerate}

\subsubsection{Evaluation using simulated gene sets and simulated data.}\label{sec:simulation_methods}

As a simple example to demonstrate the ability of the SGSE method to recover true associations between gene sets and the variance structure of genomic data, SGSE was used to compute the statistical association between 20 disjoint gene sets, each of size 10, and the spectra of 100 simulated gene expression datasets each comprised by 50 independent observations of a 200 dimension random vector simulated according to a single factor multivariate normal distribution $\sim \text{MVN}(\boldsymbol{0}, \boldsymbol{\Sigma})$. The population covariance matrix was generated as: $\boldsymbol{\Sigma} = \lambda_1 \boldsymbol{\alpha}_1 \boldsymbol{\alpha}_1^T + \lambda_d \boldsymbol{I}$, where $\lambda_1=$ 2, $\lambda_d=$ 0.1 and $\boldsymbol{\alpha}_1$ is a 200-dimensional vector with all elements equal to $0$ except for the first 10 which were set to $\sqrt{.1}$. 
The SGSE method was executed on the 100 simulated datasets using all 50 PCs with non-zero eigenvalues, default settings were used for PCGSE, as specified in Section \ref{sec:pcgse_for_sgse}, and both weighting methods outlined in Section \ref{sec:liptaks} were employed.

\subsubsection{Evaluation using MSigDB C2 v4.0 gene sets and Armstrong et al. leukemia gene expression data.}\label{sec:leukemia_methods}

The SGSE method and the benchmark cluster-based method were used to compute the statistical association between the MSigDB C2 v4.0 gene sets and the spectra of the leukemia gene expression data \citep{Armstrong:2002fk} used in the 2005 GSEA paper \citep{subramanian_gene_2005}.  The MSigDB C2 v4.0 gene sets and collapsed leukemia gene expression data were both downloaded from the MSigDB repository. With a minimum gene set size of 15 and maximum gene set size of 200, 3,076 gene sets out of the original 4,722 were used in the analysis. The SGSE method was executed on the leukemia gene expression data using all PCs with non-zero eigenvalues, PCGSE was called with default settings as specified in Section \ref{sec:pcgse_for_sgse}, and both weighting methods outlined in Section \ref{sec:liptaks} were employed. The benchmark cluster-based enrichment method was executed as outlined in Section \ref{sec:cluster_method} (k=10 was selected as optimal by the gap statistic test). The enrichment of the MSigDB C2 gene sets was also computed relative to the acute myeloid leukemia (AML) versus acute lymphoblastic leukemia (ALL) phenotype using the competitive enrichment method CAMERA \citep{Wu:2012fk} with default settings. To quantify how well SGSE and the benchmark cluster-based method were able to capture the known strong association between AML/ALL status and the second PC in the data (see analysis in \cite{Frost:PCGSE_arXiv_2014}), the Spearman correlation coefficient was calculated between unsupervised enrichment p-values and phenotype enrichment p-values for all MSigDB C2 gene sets. For gene sets with phenotype enrichment p-values less than 0.05, contingency table statistics were also computed measuring how well SGSE and the cluster-based enrichment method were able to identify MSigDB C2 gene sets significantly associated with the AML/ALL phenotype.

\subsubsection{Evaluation using Rosenwald et al. DLBCL gene expression data and MSigDB C2 v4.0 gene sets.}\label{sec:rosenwald_methods}
The SGSE method and the benchmark cluster-based method were also used to compute the statistical association between the MSigDB C2 v4.0 gene sets and the spectra of the \cite{Rosenwald:2002fk} diffuse large B-cell lymphoma (DLBCL) gene expression data. The Rosenwald et al. data set consists of gene expression measurements for 240 patients with DLBCL made using the Lymphochip microarray on 7,399 genes. Microarray data and clinical covariates from the \cite{Rosenwald:2002fk} study were both downloaded from the paper's supplemental information web site. To support spectral and phenotype enrichment analysis, the subset of the MSigDB C2 v4.0 gene sets whose members were measured in the Rosenwald et al. data was generated by mapping each of the Lymphochip probes, via Genbank accession numbers, to Entrez gene identifiers and MSigDB C2 v4.0 gene sets. Prior to execution of SGSE and the benchmark cluster-based method, all censored subjects were removed and missing values in the Rosenwald et al. data were imputed using k-nearest neighbor imputation using the impute.knn() function from the R impute package with default settings \citep{Troyanskaya:2001fk}. With a minimum gene set size of 15 and maximum gene set size of 200, 3,106 gene sets out of the original 4,722 were used in the analysis. The SGSE method was executed on the DLBCL gene expression data using all 
PCs with non-zero eigenvalues, PCGSE was called with default settings as specified in Section \ref{sec:pcgse_for_sgse}, and both weighting methods outlined in Section \ref{sec:liptaks} were employed. The benchmark cluster-based enrichment method was executed as outlined in Section \ref{sec:cluster_method} (k=10 was selected as optimal by the gap statistic test). The enrichment of the MSigDB C2 gene sets was also computed relative to the log of survival time with the competitive enrichment method CAMERA \citep{Wu:2012fk} using, as a gene-level test statistic, the z-transformed t-statistic associated with the estimated coefficient from a linear model between gene expression and log survival time. To quantify how well SGSE and the benchmark cluster-based method were able to capture the association between gene set expression and survival time, the Spearman correlation coefficient was computed between unsupervised enrichment p-values and phenotype enrichment p-values for all MSigDB C2 gene sets. For gene sets with phenotype enrichment p-values less than 0.05, contingency table statistics were computed measuring how well SGSE and the cluster-based enrichment method were able to identify MSigDB C2 gene sets significantly associated with log survival time.

\section{Results and Discussion}

\subsection{Simulation example}\label{sec:sim_results}

\begin{figure}[!ht]
\centering
\includegraphics[width=.75\textwidth]{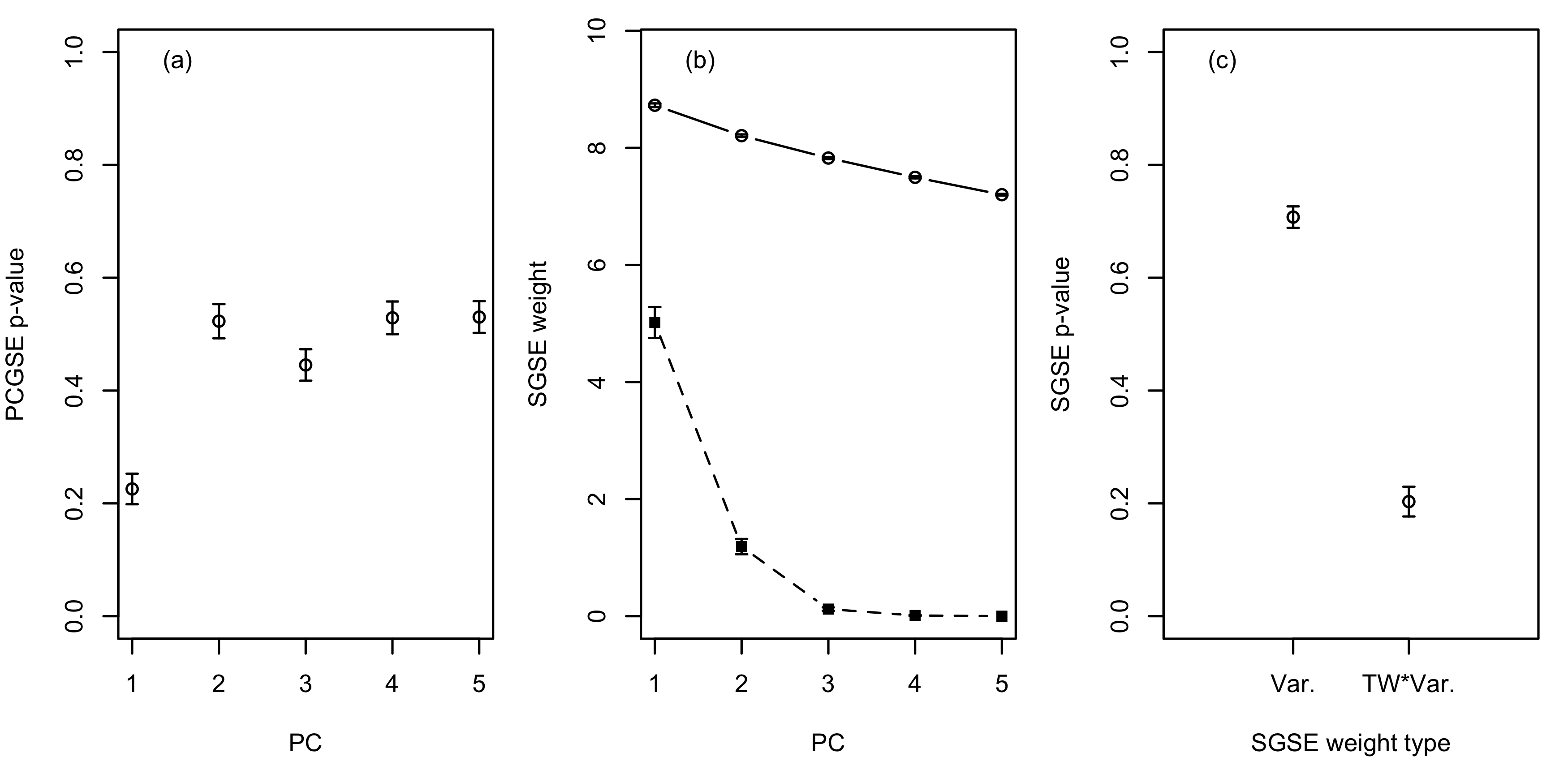}
\caption{
SGSE results on simulated single-factor data. For all plots, error bars represent $\pm 1 SE$ for the mean value over all 100 simulated datasets.
(a) Mean p-values computed using the PCGSE method for the first simulated gene set relative to the first 5 PCs over all 100 simulated datasets. 
(b) Mean weights used by the SGSE method to combine the PCGSE-computed p-values for each gene set relative to the first 5 PCs. PC variance weights are shown as round points connected by a solid line. PC variance scaled by the lower-tailed p-value computed using the \textit{Tracey-Widom} distribution for the PC variance is shown using square points connected by a dashed line.
(c) Mean p-value computed using the SGSE method for the first simulated gene set using either PC variance weights (Var.) or weights defined by the PC variance scaled by the lower-tailed \textit{Tracey-Widom} p-value of the PC variance (TW*Var.).
}
\label{fig:simulation}
\end{figure}

According to the population covariance matrix, $\boldsymbol{\Sigma}$, used to simulate the 100 datasets, only the first gene set is significantly associated with the population covariance matrix, $\boldsymbol{\Sigma}$, via the first PC. As seen in the Figure \ref{fig:simulation}, plot (a), this association is easily detected via the PCGSE method. The power of the SGSE method to detect an overall association between the first gene set and the spectra of the simulated data, however, is strongly dependent on the choice of weights used to combine the PCGSE-generated p-values in the weighted Z-method, as detailed in Section \ref{sec:liptaks}. Results from our benchmark cluster-based gene set testing method are not shown in Figure \ref{fig:simulation} because, in this case, the gap statistic only identified a single cluster in the data, making the $\chi^2$ test for gene set enrichment meaningless. Standard cluster-based methods are therefore completely uninformative in such single factor cases.

The impact of PCGSE p-value weights for the simulation example can be seen in Figure \ref{fig:simulation} plots (b) and (c). Plot (b) shows both the PC variance weights for the simulated datasets as well as weights calculated by scaling the PC variance using the lower-tailed p-value computed using the \textit{Tracey-Widom} distribution for the PC variance. This scaled PC variance weighting will result in weights being very close to the standard PC variance weights if the PC variance is highly significant according to the distribution of the principal eigenvalue of a matrix with a white Wishart distribution. As the PC variance becomes less significant, the scaling coefficient decreases lowering the effective weight for the PCGSE-computed p-value associated with that PC. As seen in Figure \ref{fig:simulation} plot (c), this significance-based scaling of the PC variance weights can have a dramatic impact on the power of the SGSE method to detect true associations that exist between groups of variables and the covariance structure of the data. For this simulation example, when the PC variance is used direct as the weight, the SGSE method is unable to identify a significant association between the first gene set and the spectra of the simulated data, however, when significance scaled PC variances are used as weights, markedly lower p-vales for the first gene set are generated.

\subsection{Leukemia gene expression example}\label{sec:leukemia_results}
\begin{center}
\begin{figure*}[!ht]
\includegraphics[width=1\textwidth]{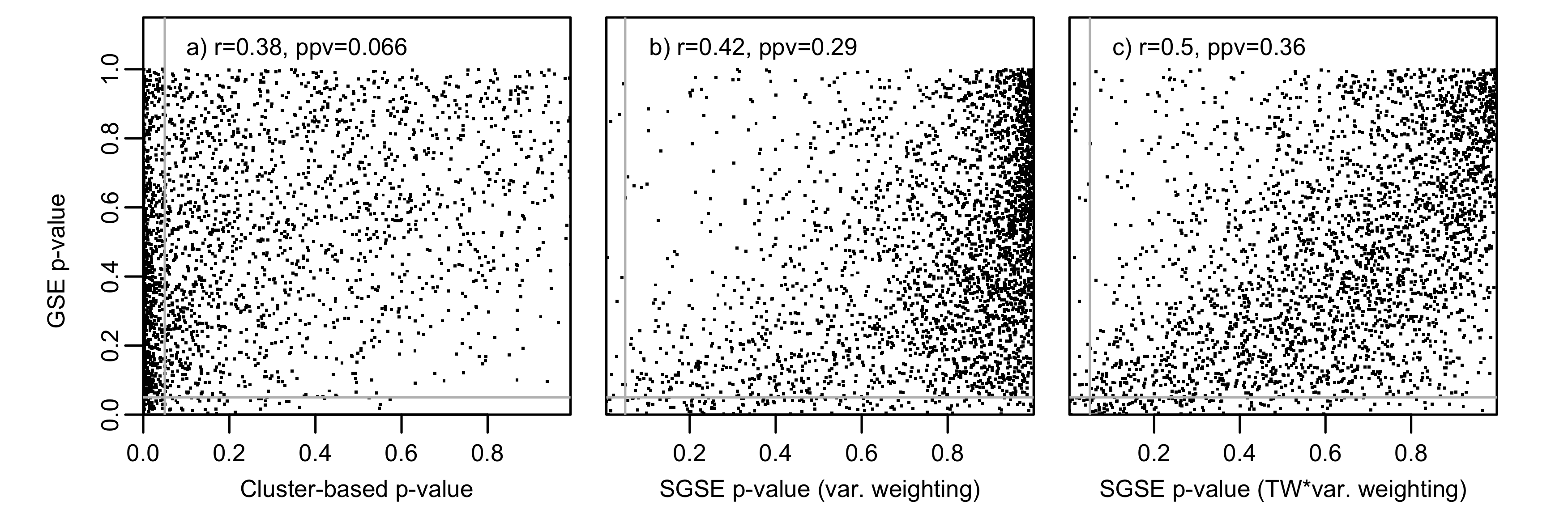}
\caption{
Scatter plot showing the association between phenotype gene set enrichment p-values and unsupervised gene set enrichment p-values computed using the benchmark cluster-based method (plot (a)) and SGSE (plots (b) and (c)) for the \cite{Armstrong:2002fk} leukemia gene expression data, AML/ALL phenotype, and MSigDB C2 v4.0 gene sets. Phenotype enrichment, unsupervised cluster-based enrichment and spectral gene set enrichment p-values were computed as outlined in Section \ref{sec:leukemia_methods}. Displayed in each plot is the Spearman correlation coefficient between phenotype and unsupervised gene set enrichment p-values and the positive predictive value of unsupervised gene set enrichment for identifying gene sets that are significantly enriched relative to the phenotype at an $\alpha=0.05$ (shown by grey lines). The results from the two different SGSE weighting methods outlined in Section \ref{sec:liptaks} are shown in plots (b) and (c) with (b) plotting SGSE p-values generated using PC variance weighting and (c) plotting SGSE p-values generated using weights defined by the PC variance scaled by the lower-tailed \textit{Tracey-Widom} p-value for the variance.
}
\label{fig:leukemia_cor}
\end{figure*}
\end{center}

The \cite{Armstrong:2002fk} leukemia gene expression dataset and MSigDB C2 v4.0 gene sets were selected for SGSE analysis because of the known association between AML/ALL status and the spectra of the gene expression data, as illustrated in \cite{Frost:PCGSE_arXiv_2014}, the easy accessibility of the data and gene sets from the MSigDB repository and the common use of both the gene expression data and curated gene sets in the gene set enrichment literature (e.g., \cite{subramanian_gene_2005}).
 
Figure \ref{fig:leukemia_cor} shows the association between phenotype and unsupervised gene set enrichment p-values computed using both the benchmark cluster-based method and the SGSE method for the MSigDB C2 v4.0 gene sets, the AML versus ALL phenotype and the Armstrong et al. leukemia gene expression data. Although the true unsupervised enrichment status of the MSigDB C2 v4.0 gene sets relative to the variance structure of the \cite{Armstrong:2002fk} gene expression data is unknown, the phenotype enrichment results can be used as a proxy for the true unsupervised gene set enrichment based on the strong association between PC 2 and AML versus ALL status \citep{Frost:PCGSE_arXiv_2014} as well as the recent finding by \cite{Gorlov:2014fk} that the genes with a large expression variance among cancer cases have a very high likelihood of having a known role in tumor-genesis. As indicated by the correlation between phenotype enrichment and unsupervised gene set enrichment p-values,  the SGSE method was able to capture a greater proportion of the AML versus ALL enrichment signal than the benchmark cluster-based method, irrespective of the method used to weight the PC-specific gene set enrichment p-values, with the best performance obtained when PC statistical significance was used to compute the SGSE weights. The benefits of the SGSE method relative to cluster-based enrichment are most clearly visible when considering identification of AML/ALL-associated gene sets via unsupervised enrichment using a phenotype enrichment threshold of $\alpha=$0.05. In this case, anti-conservative nature of the $\chi^2$ test used in the cluster-based method leads to a high type I error rate and a very low positive predictive value (PPV) of 0.066 as displayed in plot \textbf{(a)}, whereas the SGSE method has a PPV 0.29 when using PC variance weights as displayed in plot \textbf{(b)} and a PPV of 0.36 when using as weights the PC variance scaled by the lower-tailed \textit{Tracey-Widom} p-value for the variance as shown in plot \textbf{(c)}.

SGSE analysis of the MSigDB C2 v4.0 gene sets and \cite{Armstrong:2002fk} leukemia gene expression data illustrates the biological motivation for spectral gene set enrichment, shows the clear superiority of the SGSE approach relative to standard cluster-based gene set tests and demonstrates the importance of PC-specific p-value weights that take into account the statistical significance of each PC.

\subsection{DLBCL gene expression example}\label{sec:rosenwald_results}

\begin{figure*}[!ht]
\centering
\includegraphics[width=1\textwidth]{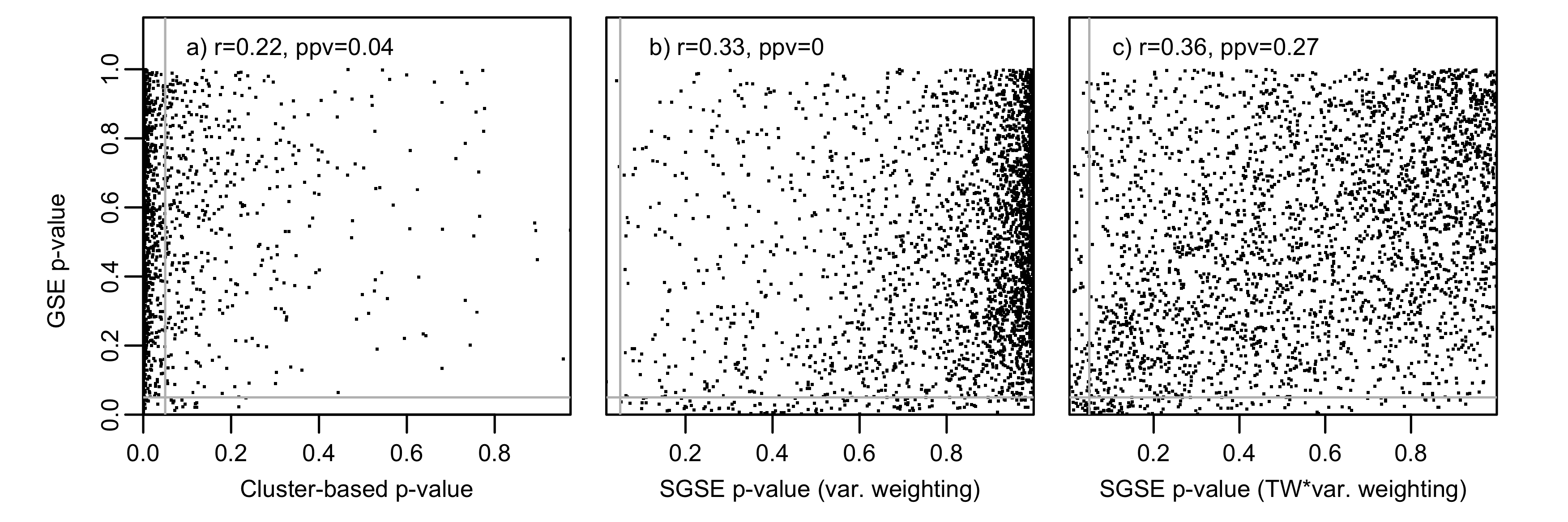}
\caption{
Scatter plot showing the association between phenotype gene set enrichment p-values and unsupervised gene set enrichment p-values computed using the benchmark cluster-based method (plot (a)) and SGSE (plots (b) and (c)) for the \cite{Rosenwald:2002fk} DLBCL gene expression data, log survival time phenotype, and MSigDB C2 v4.0 gene sets. Phenotype enrichment, unsupervised cluster-based enrichment and spectral gene set enrichment p-values were computed as outlined in Section \ref{sec:rosenwald_methods}. Displayed in each plot is the Spearman correlation coefficient between phenotype and unsupervised gene set enrichment p-values and the positive predictive value of unsupervised gene set enrichment for identifying gene sets that are significantly enriched relative to the phenotype at an $\alpha=0.05$ (shown by grey lines). The results from the two different SGSE weighting methods outlined in Section \ref{sec:liptaks} are shown in plots (b) and (c) with (b) plotting SGSE p-values generated using PC variance weighting and (c) plotting SGSE p-values generated using weights defined by the PC variance scaled by the lower-tailed \textit{Tracey-Widom} p-value for the variance.
}
\label{fig:rosenwald_cor}
\end{figure*}

The \cite{Rosenwald:2002fk} DLBCL gene expression dataset is another good example of a clear association between the variance structure of gene expression data and an interesting clinical phenotype, in this case log survival time. Similar to the Armstrong et al. leukemia gene expression data, the Rosenwald et al. DBLCL gene expression data is easily accessible and has been widely reanalyzed in the genomics literature, factors that will support interpretation and replication of the reported SGSE results by other researchers.
 
Figure \ref{fig:rosenwald_cor} shows the association between phenotype and unsupervised gene set enrichment p-values for the MSigDB C2 v4.0 gene sets, log survival time and the spectra of the Rosenwald et al. DLBCL gene expression data. Although the true enrichment status of the MSigDB C2 v4.0 gene sets relative to the variance structure of the \cite{Rosenwald:2002fk} gene expression data is unknown, the phenotype enrichment results can again be used as a proxy for the true spectral gene set enrichment based on association between expression variance and cancer-related genes \citep{Gorlov:2014fk}. Although the association between SGSE and cluster-based p-values and phenotype p-values was lower for the Rosenwald et al. DLBCL gene expression data than for the Armstrong et al. leukemia gene expression data, the SGSE method was still able to capture an appreciably greater proportion of the survival time enrichment signal as compared to the benchmark cluster-based method, irrespective of the method used to weight the PC-specific gene set enrichment p-values. Similar to the findings for the leukemia gene expression data, incorporating the PC statistical significance in the SGSE weights improved the Spearman correlation between phenotype enrichment p-values and SGSE p-values for the Rosenwald et al. data. The superior performance of the SGSE method relative to the benchmark cluster-based method was again most apparent when considering identification of survival time-associated gene sets via unsupervised enrichment using just a phenotype enrichment threshold of $\alpha=$0.05. In this case, the choice of SGSE weighting method also had a significant impact with a  positive predictive value (PPV) of 0.04 for cluster-based enrichment as displayed in plot \text{(a)}, a PPV of 0 for SGSE when using PC variance weights as displayed in plot \textbf{(b)} and a PPV of 0.27 when using as weights the PC variance scaled by the lower-tailed \textit{Tracey-Widom} p-value for the variance as shown in plot \textbf{(c)}.

\section{Conclusion}

Almost universally, gene set testing is performed in a supervised context to measure the association between functional groups of genes and a clinical phenotype. Many important examples exist, however, where a gene set-based interpretation of genomic data is desired in the absence of a  phenotype variable. Although techniques have been developed for unsupervised gene set testing, they predominantly compute enrichment relative to a categorical variable defined by disjoint clusters of the genomic variables. Because such cluster-based methods often use anti-conservative contingency table-based tests and have performance that is strongly dependent on the clustering algorithm and number of clusters, they are more useful for clustering evaluation than for gene set-based interpretation of genomic data. To address the lack of effective statistical methods for unsupervised competitive gene set testing, we have developed spectral gene set enrichment (SGSE), available in the PCGSE R package from CRAN. The SGSE method first computes the statistical association between gene sets and principal components (PCs) using our principal component gene set enrichment (PCGSE) method. The overall statistical association between each gene set and the spectral structure of the data is then computed by combining the PC-level p-values using the weighted Z-method with weights set to the PC variance scaled by lower-tailed p-values from the Tracey-Widom distribution of the eigenvalue associated with each PC.  On both simulated gene sets with simulated data and on curated gene sets with real gene expression data, the SGSE method has been shown to provide superior estimates of unsupervised gene set enrichment relative to standard cluster-based approaches.

\section*{Acknowledgement}

\paragraph{Funding:} National Institutes of Health R01 grants LM010098, LM011360, EY022300, GM103506 and GM103534.
\paragraph{Conflict of Interest:} None declared.


\bibliographystyle{natbib}
\bibliography{SGSE}

\begin{thebibliography}{}

\bibitem[Alterovitz {\em et~al.}(2010)Alterovitz, Xiang, Hill, Lomax, Liu,
  Cherkassky, Dreyfuss, Mungall, Harris, Dolan, Blake, and
  Ramoni]{alterovitz_ontology_2010}
Alterovitz, G., Xiang, M., Hill, D.~P., Lomax, J., Liu, J., Cherkassky, M.,
  Dreyfuss, J., Mungall, C., Harris, M.~A., Dolan, M.~E., Blake, J.~A., and
  Ramoni, M.~F. (2010).
\newblock Ontology engineering.
\newblock {\em Nature Biotechnology\/}, {\bf 28}(2), 128--130.

\bibitem[Armstrong {\em et~al.}(2002)Armstrong, Staunton, Silverman, Pieters,
  den Boer, Minden, Sallan, Lander, Golub, and Korsmeyer]{Armstrong:2002fk}
Armstrong, S.~A., Staunton, J.~E., Silverman, L.~B., Pieters, R., den Boer,
  M.~L., Minden, M.~D., Sallan, S.~E., Lander, E.~S., Golub, T.~R., and
  Korsmeyer, S.~J. (2002).
\newblock Mll translocations specify a distinct gene expression profile that
  distinguishes a unique leukemia.
\newblock {\em Nature Genetics\/}, {\bf 30}(1), 41--7.

\bibitem[Ashburner {\em et~al.}(2000)Ashburner, Ball, Blake, Botstein, Butler,
  Cherry, Davis, Dolinski, Dwight, Eppig, Harris, Hill, {Issel-Tarver},
  Kasarskis, Lewis, Matese, Richardson, Ringwald, Rubin, and
  Sherlock]{ashburner_gene_2000}
Ashburner, M., Ball, C.~A., Blake, J.~A., Botstein, D., Butler, H., Cherry,
  J.~M., Davis, A.~P., Dolinski, K., Dwight, S.~S., Eppig, J.~T., Harris,
  M.~A., Hill, D.~P., {Issel-Tarver}, L., Kasarskis, A., Lewis, S., Matese,
  J.~C., Richardson, J.~E., Ringwald, M., Rubin, G.~M., and Sherlock, G.
  (2000).
\newblock Gene ontology: tool for the unification of biology.
\newblock {\em Nature Genetics\/}, {\bf 25}(1), 25--29.

\bibitem[Barry {\em et~al.}(2008)Barry, Nobel, and Wright]{Barry2008}
Barry, W.~T., Nobel, A.~B., and Wright, F.~A. (2008).
\newblock A statistical framework for testing functional categories in
  microarray data.
\newblock {\em Annals of Applied Statistics\/}, {\bf 2}, 286--315+.

\bibitem[{Chiani}(2012){Chiani}]{2012arXiv1209.3394C}
{Chiani}, M. (2012).
\newblock {Distribution of the largest eigenvalue for real Wishart and Gaussian
  random matrices and a simple approximation for the Tracy-Widom distribution}.
\newblock {\em ArXiv e-prints\/}.

\bibitem[Davis {\em et~al.}(2010)Davis, Sehgal, and Ragan]{Davis:2010dq}
Davis, M.~J., Sehgal, M. S.~B., and Ragan, M.~A. (2010).
\newblock Automatic, context-specific generation of gene ontology slims.
\newblock {\em BMC Bioinformatics\/}, {\bf 11}, 498.

\bibitem[Efron and Tibshirani(2007)Efron and Tibshirani]{Efron:2007uq}
Efron, B. and Tibshirani, R. (2007).
\newblock On testing the significance of sets of genes.
\newblock {\em Annals of Applied Statistics\/}, {\bf 1}(1), 107--129.

\bibitem[Freudenberg {\em et~al.}(2009)Freudenberg, Joshi, Hu, and
  Medvedovic]{Freudenberg:2009ys}
Freudenberg, J.~M., Joshi, V.~K., Hu, Z., and Medvedovic, M. (2009).
\newblock Clean: Clustering enrichment analysis.
\newblock {\em BMC Bioinformatics\/}, {\bf 10}, 234.

\bibitem[Frost and McCray(2012)Frost and McCray]{Frost:2012zr}
Frost, H.~R. and McCray, A.~T. (2012).
\newblock Markov chain ontology analysis (mcoa).
\newblock {\em BMC Bioinformatics\/}, {\bf 13}, 23.

\bibitem[Frost and Moore(2014)Frost and Moore]{Frost:2014fk}
Frost, H.~R. and Moore, J.~H. (2014).
\newblock Optimization of gene set annotations via entropy minimization over
  variable clusters (emvc).
\newblock {\em Bioinformatics\/}.

\bibitem[{Frost} {\em et~al.}(2014){Frost}, {Li}, and
  {Moore}]{Frost:PCGSE_arXiv_2014}
{Frost}, H.~R., {Li}, Z., and {Moore}, J.~H. (2014).
\newblock {Principal component gene set enrichment (PCGSE)}.
\newblock {\em ArXiv e-prints\/}.
\newblock arXiv:1403.5148.

\bibitem[Gibbons and Roth(2002)Gibbons and Roth]{Gibbons:2002fk}
Gibbons, F.~D. and Roth, F.~P. (2002).
\newblock Judging the quality of gene expression-based clustering methods using
  gene annotation.
\newblock {\em Genome Res\/}, {\bf 12}(10), 1574--81.

\bibitem[Glaab {\em et~al.}(2012)Glaab, Baudot, Krasnogor, Schneider, and
  Valencia]{Glaab:2012kx}
Glaab, E., Baudot, A., Krasnogor, N., Schneider, R., and Valencia, A. (2012).
\newblock Enrichnet: network-based gene set enrichment analysis.
\newblock {\em Bioinformatics\/}, {\bf 28}(18), i451--i457.

\bibitem[Goeman and Buehlmann(2007)Goeman and Buehlmann]{Goeman:2007pr}
Goeman, J.~J. and Buehlmann, P. ({2007}).
\newblock {Analyzing gene expression data in terms of gene sets: methodological
  issues}.
\newblock {\em Bioinformatics\/}, {\bf {23}}({8}), {980--987}.

\bibitem[Gorlov {\em et~al.}(2014)Gorlov, Yang, Byun, Logothetis, Gorlova, Do,
  and Amos]{Gorlov:2014fk}
Gorlov, I.~P., Yang, J.-Y., Byun, J., Logothetis, C., Gorlova, O.~Y., Do,
  K.-A., and Amos, C. (2014).
\newblock How to get the most from microarray data: advice from reverse
  genomics.
\newblock {\em BMC Genomics\/}, {\bf 15}(1), 223.

\bibitem[Hartigan and Wong(1979)Hartigan and Wong]{hartigan1979}
Hartigan, J.~A. and Wong, M.~A. (1979).
\newblock {A k-means clustering algorithm}.
\newblock {\em Applied Statistics\/}, {\bf 28}(1), 100--108.

\bibitem[Hung {\em et~al.}(2012)Hung, Yang, Hu, Weng, and Delisi]{Hung:2012kx}
Hung, J.-H., Yang, T.-H., Hu, Z., Weng, Z., and Delisi, C. (2012).
\newblock Gene set enrichment analysis: performance evaluation and usage
  guidelines.
\newblock {\em Brief Bioinform\/}, {\bf 13}(3), 281--91.

\bibitem[Izenman(2008)Izenman]{izenman_rmt}
Izenman, A. (2008).
\newblock An introduction to random-matrix theory.

\bibitem[Johnstone(2001)Johnstone]{johnstone_2001}
Johnstone, I.~M. (2001).
\newblock On the distribution of the largest eigenvalue in principal components
  analysis.
\newblock {\em The Annals of Statistics\/}, {\bf 29}(2), pp. 295--327.

\bibitem[Johnstone(2009)Johnstone]{Johnstone:2009fk}
Johnstone, I.~M. (2009).
\newblock Approximate null distribution of the largest root in multivariate
  analysis.
\newblock {\em Ann Appl Stat\/}, {\bf 3}(4), 1616--1633.

\bibitem[Jolliffe(2002)Jolliffe]{jolliffe2002principal}
Jolliffe, I. (2002).
\newblock {\em Principal Component Analysis\/}.
\newblock Springer Series in Statistics. Springer.

\bibitem[Kanehisa and Goto(2000)Kanehisa and Goto]{kanehisa_kegg:_2000}
Kanehisa, M. and Goto, S. (2000).
\newblock {KEGG:} kyoto encyclopedia of genes and genomes.
\newblock {\em Nucleic Acids Research\/}, {\bf 28}(1), 27--30.

\bibitem[Kaufman and Rousseeuw(2005)Kaufman and Rousseeuw]{Kaufman:2005vn}
Kaufman, L. and Rousseeuw, P.~J. (2005).
\newblock {\em Finding groups in data: an introduction to cluster analysis\/}.
\newblock Wiley, Hoboken, N.J.

\bibitem[Khatri {\em et~al.}(2012)Khatri, Sirota, and Butte]{Khatri:2012fk}
Khatri, P., Sirota, M., and Butte, A.~J. (2012).
\newblock Ten years of pathway analysis: current approaches and outstanding
  challenges.
\newblock {\em {PLoS} Computational Biology\/}, {\bf 8}(2), e1002375.

\bibitem[Kost and McDermott(2002)Kost and McDermott]{Kost2002183}
Kost, J.~T. and McDermott, M.~P. (2002).
\newblock Combining dependent p-values.
\newblock {\em Statistics and Probability Letters\/}, {\bf 60}(2), 183 -- 190.

\bibitem[Lee and Batzoglou(2003)Lee and Batzoglou]{Lee:2003bh}
Lee, S.-I. and Batzoglou, S. (2003).
\newblock Application of independent component analysis to microarrays.
\newblock {\em Genome Biol\/}, {\bf 4}(11), R76.

\bibitem[Liberzon {\em et~al.}(2011)Liberzon, Subramanian, Pinchback,
  Thorvaldsd{\'o}ttir, Tamayo, and Mesirov]{Liberzon:2011uq}
Liberzon, A., Subramanian, A., Pinchback, R., Thorvaldsd{\'o}ttir, H., Tamayo,
  P., and Mesirov, J.~P. (2011).
\newblock Molecular signatures database (msigdb) 3.0.
\newblock {\em Bioinformatics\/}, {\bf 27}(12), 1739--40.

\bibitem[Maechler {\em et~al.}(2014)Maechler, Rousseeuw, Struyf, Hubert, and
  Hornik]{Maechler:2014fk}
Maechler, M., Rousseeuw, P., Struyf, A., Hubert, M., and Hornik, K. (2014).
\newblock {\em cluster: Cluster Analysis Basics and Extensions\/}.
\newblock R package version 1.15.2 --- For new features, see the 'Changelog'
  file (in the package source).

\bibitem[Patterson {\em et~al.}(2006)Patterson, Price, and
  Reich]{Patterson:2006uq}
Patterson, N., Price, A.~L., and Reich, D. (2006).
\newblock Population structure and eigenanalysis.
\newblock {\em PLOS Genetics\/}, {\bf 2}(12), e190.

\bibitem[Ramsay {\em et~al.}(1984)Ramsay, Berge, and Styan]{Ramsay:1984fk}
Ramsay, J., Berge, J., and Styan, G. (1984).
\newblock Matrix correlation.
\newblock {\em Psychometrika\/}, {\bf 49}, 403--423.

\bibitem[Robinson {\em et~al.}(2002)Robinson, Grigull, Mohammad, and
  Hughes]{Robinson:2002fk}
Robinson, M.~D., Grigull, J., Mohammad, N., and Hughes, T.~R. (2002).
\newblock Funspec: a web-based cluster interpreter for yeast.
\newblock {\em BMC Bioinformatics\/}, {\bf 3}, 35.

\bibitem[Roden {\em et~al.}(2006)Roden, King, Trout, Mortazavi, Wold, and
  Hart]{roden_mining_2006}
Roden, J.~C., King, B.~W., Trout, D., Mortazavi, A., Wold, B.~J., and Hart,
  C.~E. (2006).
\newblock Mining gene expression data by interpreting principal components.
\newblock {\em {BMC} Bioinformatics\/}, {\bf 7}, 194.

\bibitem[Rosenwald {\em et~al.}(2002)Rosenwald, Wright, Chan, Connors, Campo,
  Fisher, Gascoyne, Muller-Hermelink, Smeland, Giltnane, Hurt, Zhao, Averett,
  Yang, Wilson, Jaffe, Simon, Klausner, Powell, Duffey, Longo, Greiner,
  Weisenburger, Sanger, Dave, Lynch, Vose, Armitage, Montserrat,
  L{\'o}pez-Guillermo, Grogan, Miller, LeBlanc, Ott, Kvaloy, Delabie, Holte,
  Krajci, Stokke, Staudt, and {Lymphoma/Leukemia Molecular Profiling
  Project}]{Rosenwald:2002fk}
Rosenwald, A., Wright, G., Chan, W.~C., Connors, J.~M., Campo, E., Fisher,
  R.~I., Gascoyne, R.~D., Muller-Hermelink, H.~K., Smeland, E.~B., Giltnane,
  J.~M., Hurt, E.~M., Zhao, H., Averett, L., Yang, L., Wilson, W.~H., Jaffe,
  E.~S., Simon, R., Klausner, R.~D., Powell, J., Duffey, P.~L., Longo, D.~L.,
  Greiner, T.~C., Weisenburger, D.~D., Sanger, W.~G., Dave, B.~J., Lynch,
  J.~C., Vose, J., Armitage, J.~O., Montserrat, E., L{\'o}pez-Guillermo, A.,
  Grogan, T.~M., Miller, T.~P., LeBlanc, M., Ott, G., Kvaloy, S., Delabie, J.,
  Holte, H., Krajci, P., Stokke, T., Staudt, L.~M., and {Lymphoma/Leukemia
  Molecular Profiling Project} (2002).
\newblock The use of molecular profiling to predict survival after chemotherapy
  for diffuse large-b-cell lymphoma.
\newblock {\em N Engl J Med\/}, {\bf 346}(25), 1937--47.

\bibitem[Soshnikov(2002)Soshnikov]{Soshnikov02anote}
Soshnikov, A. (2002).
\newblock A note on universality of the distribution of the largest eigenvalues
  in certain sample covariance matrices.
\newblock {\em J. Statist. Phys\/}, {\bf 108}, 1033--1056.

\bibitem[Steuer {\em et~al.}(2006)Steuer, Humburg, and Selbig]{Steuer:2006vn}
Steuer, R., Humburg, P., and Selbig, J. (2006).
\newblock Validation and functional annotation of expression-based clusters
  based on gene ontology.
\newblock {\em BMC Bioinformatics\/}, {\bf 7}, 380.

\bibitem[Subramanian {\em et~al.}(2005)Subramanian, Tamayo, Mootha, Mukherjee,
  Ebert, Gillette, Paulovich, Pomeroy, Golub, Lander, and
  Mesirov]{subramanian_gene_2005}
Subramanian, A., Tamayo, P., Mootha, V.~K., Mukherjee, S., Ebert, B.~L.,
  Gillette, M.~A., Paulovich, A., Pomeroy, S.~L., Golub, T.~R., Lander, E.~S.,
  and Mesirov, J.~P. (2005).
\newblock Gene set enrichment analysis: A knowledge-based approach for
  interpreting genome-wide expression profiles.
\newblock {\em Proc Natl Acad Sci U S A\/}, {\bf 102}(43), 15545--15550.

\bibitem[Tian {\em et~al.}(2005)Tian, Greenberg, Kong, Altschuler, Kohane, and
  Park]{Tian:2005zr}
Tian, L., Greenberg, S.~A., Kong, S.~W., Altschuler, J., Kohane, I.~S., and
  Park, P.~J. (2005).
\newblock Discovering statistically significant pathways in expression
  profiling studies.
\newblock {\em Proc Natl Acad Sci U S A\/}, {\bf 102}(38), 13544--9.

\bibitem[Tibshirani {\em et~al.}(2001)Tibshirani, Walther, and
  Hastie]{Tibshirani:2001fk}
Tibshirani, R., Walther, G., and Hastie, T. ({2001}).
\newblock {Estimating the number of clusters in a data set via the gap
  statistic}.
\newblock {\em {Journal of the Royal Statistical Society. Series B
  (Methodological)}\/}, {\bf {63}}({Part 2}), {411--423}.

\bibitem[Toronen(2004)Toronen]{Toronen:2004uq}
Toronen, P. (2004).
\newblock Selection of informative clusters from hierarchical cluster tree with
  gene classes.
\newblock {\em BMC Bioinformatics\/}, {\bf 5}, 32.

\bibitem[Tracy and Widom(1994)Tracy and Widom]{Tracy:1994fk}
Tracy, C. and Widom, H. (1994).
\newblock Level-spacing distributions and the airy kernel.
\newblock {\em Communications in Mathematical Physics\/}, {\bf 159}(1),
  151--174.

\bibitem[Troyanskaya {\em et~al.}(2001)Troyanskaya, Cantor, Sherlock, Brown,
  Hastie, Tibshirani, Botstein, and Altman]{Troyanskaya:2001fk}
Troyanskaya, O., Cantor, M., Sherlock, G., Brown, P., Hastie, T., Tibshirani,
  R., Botstein, D., and Altman, R.~B. (2001).
\newblock Missing value estimation methods for dna microarrays.
\newblock {\em Bioinformatics\/}, {\bf 17}(6), 520--5.

\bibitem[Whitlock(2005)Whitlock]{Whitlock:2005uq}
Whitlock, M.~C. (2005).
\newblock Combining probability from independent tests: the weighted z-method
  is superior to fisher's approach.
\newblock {\em J Evol Biol\/}, {\bf 18}(5), 1368--73.

\bibitem[Wolfe {\em et~al.}(2005)Wolfe, Kohane, and Butte]{Wolfe:2005ly}
Wolfe, C.~J., Kohane, I.~S., and Butte, A.~J. (2005).
\newblock Systematic survey reveals general applicability of
  "guilt-by-association" within gene coexpression networks.
\newblock {\em BMC Bioinformatics\/}, {\bf 6}, 227.

\bibitem[Won {\em et~al.}(2009)Won, Morris, Lu, and Elston]{Won:2009fk}
Won, S., Morris, N., Lu, Q., and Elston, R.~C. (2009).
\newblock Choosing an optimal method to combine p-values.
\newblock {\em Stat Med\/}, {\bf 28}(11), 1537--53.

\bibitem[Wu and Smyth(2012)Wu and Smyth]{Wu:2012fk}
Wu, D. and Smyth, G.~K. (2012).
\newblock Camera: a competitive gene set test accounting for inter-gene
  correlation.
\newblock {\em Nucleic Acids Research\/}, {\bf 40}(17), e133.

\bibitem[Yao {\em et~al.}(2012)Yao, Coquery, and L{\^e}~Cao]{Yao:2012qf}
Yao, F., Coquery, J., and L{\^e}~Cao, K.-A. (2012).
\newblock Independent principal component analysis for biologically meaningful
  dimension reduction of large biological data sets.
\newblock {\em BMC Bioinformatics\/}, {\bf 13}, 24.

\bibitem[Zhao {\em et~al.}(2010)Zhao, Langfelder, Fuller, Dong, Li, and
  Hovarth]{Zhao:2010fk}
Zhao, W., Langfelder, P., Fuller, T., Dong, J., Li, A., and Hovarth, S. (2010).
\newblock Weighted gene coexpression network analysis: state of the art.
\newblock {\em J Biopharm Stat\/}, {\bf 20}(2), 281--300.

\bibitem[Zhou {\em et~al.}(2013)Zhou, Barry, and Wright]{Zhou:2013ys}
Zhou, Y.-H., Barry, W.~T., and Wright, F.~A. (2013).
\newblock Empirical pathway analysis, without permutation.
\newblock {\em Biostatistics\/}, {\bf 14}(3), 573--85.

\end{thebibliography}

\end{document}